\RequirePackage{fix-cm}
\documentclass[smallextended]{svjour3}       

\smartqed  
\usepackage{graphicx}
\usepackage{marvosym}
\usepackage{color}
\usepackage{amsmath}
\usepackage{soul}
\usepackage{amsfonts}
\usepackage{verbatim}
\begin{document}
\title{A Critical and Moving-Forward View on Quantum Image Processing
}


\author{Fei Yan \and
        Salvador E. Venegas-Andraca \and
        Kaoru Hirota
}


\institute{F. Yan \at
              School of Computer Science and Technology, Changchun University of Science and Technology, China
           \and
           S.E. Venegas-Andraca \Letter\at
              Escuela de Ingenieria y Ciencias, Tecnologico de Monterrey, Mexico\\
              \email{salvador.venegas-andraca@keble.oxon.org}
           \and
           K. Hirota \at
              Department of Computational Intelligence and Systems Science, Tokyo Institute of Technology, Japan
}

\date{Received: date / Accepted: date}

\maketitle


\vspace{0.1cm}

Physics and computer science have a long tradition of cross-fertilization. One of the latest outcomes of this mutually beneficial relationship is quantum information science, which comprises the study of information processing tasks that can be accomplished using quantum mechanical systems \cite{2}.  Quantum Image Processing (QIMP) is an emergent field of quantum information science whose main goal is to strengthen our capacity for storing, processing, and retrieving visual information from images and video either by transitioning from digital to quantum paradigms or by complementing digital imaging with quantum techniques. The expectation is that harnessing the properties of quantum mechanical systems in QIMP will result in the realization of advanced technologies that will outperform, enhance or complement existing and upcoming digital technologies for image and video processing tasks.


QIMP has become a popular area of quantum research due to the ubiquity and primacy of digital image and video processing in modern life \cite{3}. Digital image processing is a key component of several  branches of applied computer science and engineering like computer vision and pattern recognition, disciplines that have had a tremendous scientific, technological and commercial success due to their widespread applications in many fields like medicine \cite{3a,3b}, military technology \cite{3bb,3bbb} and the entertaining industry \cite{3c,3d}. The technological and commercial success of digital image processing in contemporary (both civil and military) life is a most powerful incentive for working on QIMP.


A key feature of QIMP which is crucial to understand its current development and challenges, as well as to design corresponding science and technology roadmaps, is the following: QIMP is both a scientific discipline and a field of engineering with an eye on developing commercial applications. Potential applications of QIMP are to be found not only in the development of quantum algorithms for general-purpose quantum computers but also in specific-purpose technology like quantum radar \cite{3dd} and smart cameras, for instance.

So, there is a plethora of reasons to work in QIMP: some motivations could be of scientific theoretical nature while others could be focused on integrating classical and quantum technologies aiming at developing products and services for high-tech markets. Thus, research interests in QIMP are diverse, some will fall within the traditional scope of quantum computing scientific research (like using quantum entanglement for information processing \cite{3e}, quantum mathematical morphology \cite{3f}, image segmentation \cite{3g} or designing quantum algorithms with provable computational speed-up, being this last topic a pending task in the QIMP community) while other approaches will be oriented towards engineering applications (e.g., \cite{3h,3i}).


Since QIMP was proposed by Venegas-Andraca and Bose in 2003 \cite{4ijcai,4,5}, the volume of research focused on QIMP has steadily increased in terms of the number of papers published each year in China and other countries, as shown in Fig. \ref{fig1} (QIMP is of particular interest to Chinese research groups, hence China has become the most important contributor to this discipline nowadays.) Fig. \ref{fig1} also presents a succinct analysis of QIMP subtopics and the corresponding percentages of published papers according to the Web of Science database. We note that the vast majority of QIMP research has been devoted to security issues (41.9\%), while the least work has focused on quantum image representation (8.9\%). This analysis of present developments in the field also highlights several features and further research topics in QIMP.

This paper seeks to arouse the interest of scientific and engineering communities towards the greater realization of QIMP-based technologies by identifying and discussing three primary issues whose development would be most helpful for advancing the field of QIMP. Before addressing these issues, we would like to present the following analysis:

\begin{itemize}
\item[{\tiny \textbullet}] The original motivations to create QIMP as presented in \cite{4ijcai,4,5} are similar to those that gave birth to the field of quantum walks \cite{5c} in the sense that both disciplines were born  as quantum counterparts of already existing mathematical and computational tools in the digital world (that we can tell because one of authors of \cite{4ijcai,4,5} is also an author of this perspective paper).

First steps in QIMP were focused on showing that it was indeed possible to create a set of tools in the quantum domain, somewhat equivalent to those already developed in the digital domain, that would provide us with basic capacities for storing, manipulating and retrieving images stored in quantum systems. Indeed, for any classical computation there is a quantum counterpart (i.e., classical computation is a subset of quantum computation) but in order to convince scientists and engineers from fields other than quantum computing, whose academic background would not necessarily include quantum mechanics, it was compulsory to present solid evidence of the viability of QIMP, evidence written and described in the language and techniques of digital image processing.

\begin{figure*}
   (a) \includegraphics[width=.58\textwidth]{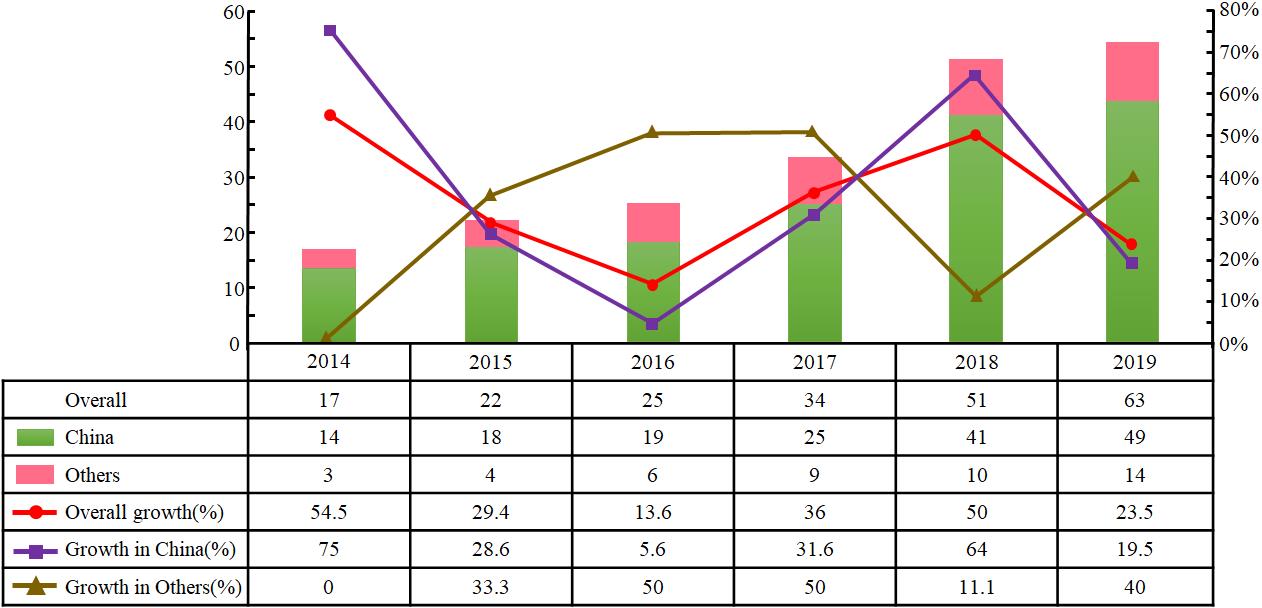}\hfill
   (b) \includegraphics[width=.28\textwidth]{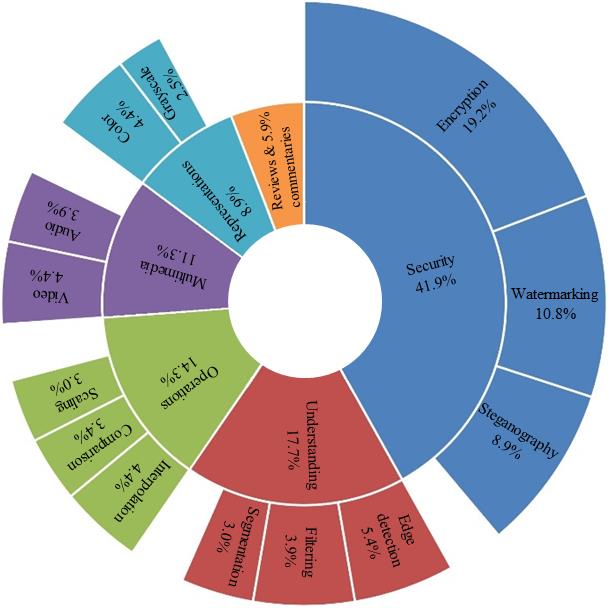}\hfill
\caption{(a) Bar graph and corresponding table data showing the number of papers published in the field of QIMP among researchers in China and other countries since 2014 and the percentile change in the number of papers published each year relative to the previous year. (b) Sunburst chart providing a rough summary and classification of available results in the QIMP field, which demonstrates the vigorous and wide-ranging nature of QIMP research over the past five years. It is worthy of note that the statistics above considered only publications included in the Web of Science database obtained from a search based on the keywords ``quantum image processing''. Moreover, studies that merely cited QIMP but did not contribute to the field were not counted.}
\label{fig1}
\end{figure*}

So, the QIMP community worked on designing methods that would {\it in principle} allow us to encode, process and recover images using quantum systems \cite{7}, followed by algorithms that would provide us with key routines and capacities like arbitrary rotations, scaling, similarity evaluation, encryption and steganography \cite{3}. The focus was on making sure that algorithms were robust rather than efficient (or more efficient than their classical counterparts.) A full introduction to QIMP can be found in \cite{3} and concise reviews on quantum image representation models and security technologies have been published in \cite{5b,7}.

\item[{\tiny \textbullet}] The historical development of QIMP is in close resemblance to the early evolution of digital image processing back in the period 1950-1971 \cite{5d,5e} as, for more than two decades, researchers worked on building a theoretical corpus for the storage and basic operations on digital images (as stated in \cite{5d}, \lq \lq Over the past 15 years, much effort has been devoted to developing methods of processing pictorial information by computer.'') Moreover, note that although available computer power at that time was not enough to compute beyond some simple tasks, that  did not prevent scientists from working on the algorithms that would eventually be run by the end of the century. In order to appreciate the advances of QIMP and rightly situate its challenges, an exercise of comparative history is both necessary and helpful (in fact, this exercise of comparative history should also be made for quantum algorithms and other branches of quantum computing.)

\item[{\tiny \textbullet}] From its inception, QIMP has benefited from the talent and efforts of a research community vastly composed of computer scientists, mathematicians and computer engineers. Now, as all branches of quantum computing, unleashing the power of QIMP depends upon a full interdisciplinary approach in which physicists, chemists and other professional communities also actively contribute towards solving open problems and challenges in this field (for instance, the development of novel quantum image representation models and applications, as proposed in the following lines.) This approach is  particularly important to QIMP because of the existence of Quantum Imaging \cite{5eeqimaging01,5eeqimaging02,5eeqimaging03}, a branch of quantum optics and quantum information focused on harnessing quantum correlations and other properties of quantum mechanical systems in order to surmount imaging limits imposed by classical optics (e.g., \cite{5eeqimaging04}). Coordinating research efforts from QIMP and Quantum Imaging communities towards common goals would certainly boost research and investment in scientific research and innovative applications.

 \end{itemize}

Based on the above analysis, we now discuss the three issues mentioned earlier: 1. Storage and retrieval of images in quantum systems; 2. Algorithm development and algorithmic speed-up in QIMP; and 3. The road ahead - a proposal of future steps for QIMP.
\vspace{0.3cm}

\begin{figure*}
   \includegraphics[width=1\textwidth]{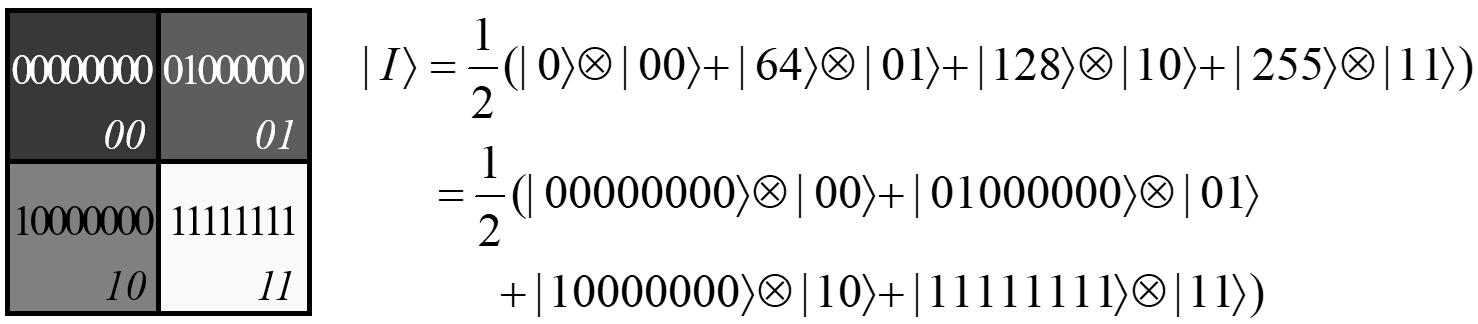}
\caption{(a) A $2\times2$ NEQR image and its quantum state.}
\label{fig2-38}
\end{figure*}

{\bf 1. Storage and retrieval of images in quantum systems}. Digital images are discrete representations of the physical world, they are stored in arrays resembling matrices $M$ of order $m_1 \times m_2$ and each entry $a_{i,j} \in M $ is a picture element, a.k.a. pixel, produced via the photoelectric effect \cite{5f}. Pixels are arrays of bits, usually $8$ bits for grayscale images and $24$ bits for color images on the RGB model (this is the standard in commercial imaging, cameras designed for scientific research  may use more bits, depending on specific needs.) So, pixels are scalars, the total number of bits required to store a digital image as described in this paragraph is $8 \times m_1 \times m_2$ for grayscale images and $24 \times m_1 \times m_2$ for RGB color images, and operations on digital images are usually described as operations on matrices or pixel-wise functions.

In QIMP, the term {\it quantum image} was coined to refer to an image stored in a set of qubits, regardless of the classical/quantum nature of the source of information, and operations on quantum images are performed via quantum evolution \cite{7}. Several methods for creating quantum images (known as image representation models) have been proposed,  being the Flexible Representation for Quantum Images (FRQI) \cite{6} and the Novel Enhanced Quantum Representation (NEQR) \cite{zhang2013neqr} models particularly popular among them. Let us analyze the NEQR representation model in order to quantify the amount of resources needed to store and retrieve information contained in quantum images.

Let us suppose we have a spatially-ordered array $A$ of $m^2$ colors (i.e., frequencies) denoted by $\{  \theta_0, \theta_1, \ldots \theta_{m^2 -1} \}$. We want to store them using two formats: a digital image and a quantum image. Let us assume, for the sake of convenience (i.e., just to avoid cumbersome calculations) and without loss of generality, that $m = 2^r$, where $r \in \mathbb{N}$. It is worth noting that the size of array $A$, $m = 2^r$, {\it is not meant to asymptotically grow exponentially large as $m$ is a fixed number, not a function}. Modern digital images produced by professional cameras are of the order of megapixels. For instance, the $\alpha 9^{\tiny{\textregistered}}$ Sony$^{\tiny{\textregistered}}$ camera has a 24.2 megapixel resolution, that is $2.42 \times 10^7$ pixels per digital photograph. Thus, $m^2 = 2.42 \times 10^7 \Rightarrow m \approx 4920$. Since $2^{12} = 4096$ and $ 2^{13} = 8192$, we may set $r= 13$ as an upper bound for modern digital camera technology.


So, to store $A$ as a digital image, we would need $m^2$ pixels and $8$ bits for each grayscale pixel or $24$ bits for each color pixel in the RGB model, i.e., $24m^2$ bits in total at most. Note that we are only counting the number of bits needed to store an $m^2$ digital RGB color image, we are not taking into account the hardware and processing power required by the Nyquist-Shannon sampling theorem to avoid aliasing and excessive blur.
\newline{}


Let us now focus on resource consumption on quantum images. Eq. (\ref{neqr1}) introduces the $m \times m$ NEQR image model.

\begin{equation}\label{neqr1}
\vert I\rangle = \frac{1}{m}\sum_{y=0}^{m-1}\sum_{x=0}^{m-1}\vert f(y,x)\rangle\vert yx\rangle\\
=\frac{1}{m}\sum_{y=0}^{m-1}\sum_{x=0}^{m-1}\bigotimes_{i=0}^{q-1}\vert C_{yx}^i\rangle\vert yx\rangle
\end{equation}
where $f(y,x)$ is a grayscale value taken from the set $\{0, 1, \ldots, 2^q-1\}$ and written as $f(y,x) = C_{yx}^{q-1}C_{yx}^{q-2}\dots C_{yx}^1C_{yx}^0$, where $C_{yx}^i\in\{0,1\}$.  Quantum states $|C_{yx}^i\rangle$ used to store grayscale values in the NEQR model are elements of the computational basis of ${\cal H}^{2^q}$ \cite{7}. 



For instance, storing in the NEQR model an image composed of $m^2$ pixels, being the value of each pixel a grasycale value taken from the set $\{0, 1, \ldots 255\}$,  would translate Eq.(\ref{neqr1}) into Eq.(\ref{neqr2})

\begin{equation}\label{neqr2}
\vert I\rangle = \frac{1}{m}\sum_{y=0}^{m-1}\sum_{x=0}^{m-1}\vert f(y,x)\rangle\vert yx\rangle\\
=\frac{1}{m}\sum_{y=0}^{m-1}\sum_{x=0}^{m-1}\bigotimes_{i=0}^{7}\vert C_{yx}^i\rangle\vert yx\rangle
\end{equation}

where quantum states $|C_{yx}^i\rangle$ are elements of the computational basis of ${\cal H}^8$.

Storing an $m^2$ pixel  image with pixel values taken from $\{0,1, \ldots, 255\}$ grayscales  on the NEQR model takes $ 8m^2 + \log_2 m^2$ qubits: $8m^2$ qubits to store $m^2$ pixels whose grayscale value is taken from  $\{0, 1, \ldots , 255\}$ and $\log_2 m^2$ qubits as indices. An example of a $2\times2$ NEQR image is shown in Fig.~\ref{fig2-38}.

\newpage{}

So, in a digital system, it takes $24 m^2$ bits to store a color image and $8 m^2$ bits to store a grayscale image, while  $ 8m^2 + \log_2 m^2$ qubits are needed to store a grayscale image in the NEQR model, i.e. in both models we need a polynomial amount of resources to store a color/grayscale image. In this regard,  there is room for improvement as qubits are scarce and it is hard to justify using quantum states as indices while this is a job that could be done using bits. This criticism points towards the development of hybrid classical-quantum approaches for storing, processing and retrieving quantum images.

Now we address a most important criticism that has been raised about quantum image representation models. In general (except for the NEQR model in which grayscale values can be deterministically retrieved, as it suffices to individually measure quantum states $|C_{yx}^i\rangle$ with measurement operators based on the computational basis of ${\cal H}^{2^q}$), retrieving an image stored in a quantum image representation model is a probabilistic process that would require identical preparation of several quantum images. In contrast, retrieving a color image stored in a digital system is a deterministic process that requires only one copy. Thus, the amount of resources required {\it to retrieve} images stored in quantum systems is larger than in the digital case (this resource estimation analysis cannot be forthrightly extended to the full procedure of acquiring, processing and measuring quantum images because, on the one hand, in this paper we have not taken into account the resources needed to implement the Nyquist-Shannon sampling theorem on digital images and, on the other hand, quantum hardware for QIMP has not been designed yet).


Indeed, under most current quantum image representation models, image retrieval is a probabilistic process. This is also the case with all other branches of quantum computation, including most celebrated quantum algorithms like Shor's and Grover's \cite{2}: data is extracted from a quantum system via a probabilistic procedure. In this sense, the criticism is right but it must be contextualized as this is a feature shared with other fields of quantum computation because it is inherited from quantum mechanics. So, a crucial research goal in QIMP must be to design algorithms and measurement strategies that reduce the amount of quantum resources required for data manipulation and data extraction. In the following lines, we elaborate some forward-looking ideas on quantum image retrieval and quantum image representation models.

\begin{itemize}

 \item[{\tiny $\bullet$}] Quantum image representation models developed so far are mostly inspired in digital image formats and do not fully incorporate quantum mechanical properties. Current models of quantum image representation require extensive improvement to integrate and take full advantage of quantum mechanical properties as well as to design efficient strategies for image retrieval. A good example of the revolutionary potential of quantum image representations lies in the use of quantum entanglement as a resource to natively store depth information in a quantum image, and thereby overcoming the geometrical constraints imposed by the $\mathbb{R}^3\rightarrow\mathbb{P}^2$ transformation that governs the geometry of digital images.

\newpage{}
Next generations of quantum image models must be fully-fledged (possibly hybrid in the sense described above) methods for storing visual information with quantum-mechanical properties as essential components (for example, \cite{grigoryan2020} is a promising approach along these lines). This is key in order to integrate quantum images as a full member of quantum technology ecosystems like quantum radar technology and other novel applications.

 \item[{\tiny $\bullet$}] So far, techniques used in QIMP for image retrieval consist of basic projective measurements and straightforward use of statistics, leaving aside the full richness of quantum measurement theory. Next steps in QIMP research programs must include quantum state tomography  techniques \cite{7a} that, possibly combined with advanced computational paradigms like machine learning \cite{7aa}  would potentially lead to optimized quantum image retrieval processes.

\item[{\tiny $\bullet$}] A potential fruitful area for further development of QIMP is to test upcoming quantum image representation models, algorithms and information retrieval techniques in fields  in which images are likey to play a key role. Let us now mention a few examples along this line of thought:

\begin{itemize}

\item The emergence and consolidation of quantum technology as a pervasive field in science, engineering and other disciplines will come together with a dire necessity to keep data safe, which translates into the creation of encryption and steganography methods for quantum images (some recent examples along these lines are \cite{7aa1,7aa2,7d}).

According to Fig. \ref{fig1}, the area of security on quantum images is most popular among QIMP researchers, hence we would have the human capital needed to design, for example, homomorphic encryption techniques  for quantum images that, among other novel methods for encrypting quantum information, could provide us with novel levels of secrecy (see \cite{7aa3}  for a recent review on homomorphic encryption).

\item Images are a key component in advanced fields of classical computer science and engineering like computer vision \cite{7aa4}, artificial intelligence \cite{7aa5}, pattern recognition and machine learning  \cite{7aa6}. A promising research avenue would be to design quantum image storage and retrieval methods suitable for being used as input to algorithms from the emergent fields of quantum machine learning \cite{8,9,10}, quantum computer vision (e.g., \cite{7aa7}) and quantum pattern recognition \cite{7aa8}.  Three concrete examples would be quantum algorithms for reconstructing medical images \cite{kiani2020}, variational quantum algorithms \cite{7aa9,7aa10,7aa11}  for quantum machine learning and other emergent branches of quantum computing, as well as  quantum clustering algorithms \cite{7aa12,7aa13,7aa14}, a promising area of unsupervised quantum machine learning whose classical counterparts are most useful in pattern recognition (for introductions to classical clustering algorithms, see \cite{7aa15,7aa16}).

\end{itemize}

\end{itemize}

\newpage{}
{\bf 2. Algorithm development and algorithmic speed-up in QIMP}.  As shown in Fig. \ref{fig1}, quantum algorithms in QIMP have been developed in six main areas, being security the most popular among them. A review of the papers published in the areas of image representation, operations, multimedia and understanding reveals that the efforts of the QIMP community have been focused on developing techniques that would replicate or complement existing techniques in the digital image processing domain, being the main goal that of algorithm robustness (that is, that the algorithm does what it is meant and expected to do) rather than speed-up.

As previously stated in this paper, arguments on computational complexity must be contextualized in order to make fair comparisons. To date, just a few cases of algorithmic speed-up are known in the quantum computing domain. Grover's algorithm is irrefutably faster than its best possible classical counterpart (so, Grover's is provably faster and optimal). The other most celebrated quantum algorithm, Shor's algorithm, is exponentially faster than the best {\it known} classical counterparts designed so far (i.e., we know that factorization is in $\mathsf{NP}$ but we do not know whether factorization is in $\mathsf{P}$ or not). A third case is presented in \cite{childsqw}, where a quantum walk-based algorithm running on glued-tress (a very particular family of trees designed specifically for this algorithm) is exponentially faster than its classical random walk-based counterpart. Both in classical and quantum computing, achieving exponential speed-up is a very difficult task, most of the quantum algorithms developed so far that are faster than their classical counterparts exhibit quadratic speed-up.

To date and to the best of our knowledge, there is no QIMP algorithm that exhibits irrefutable exponential speed-up with respect to a classical counterpart, i.e., a classical image processing algorithm. In the following lines, we argue that imposing on QIMP the requirement of exponential speed-up as the main and/or only success criterion is a choice of limited scope that does not take into account the nature and goals of this discipline.



\begin{figure}
   (a) \includegraphics[width=.28\textwidth]{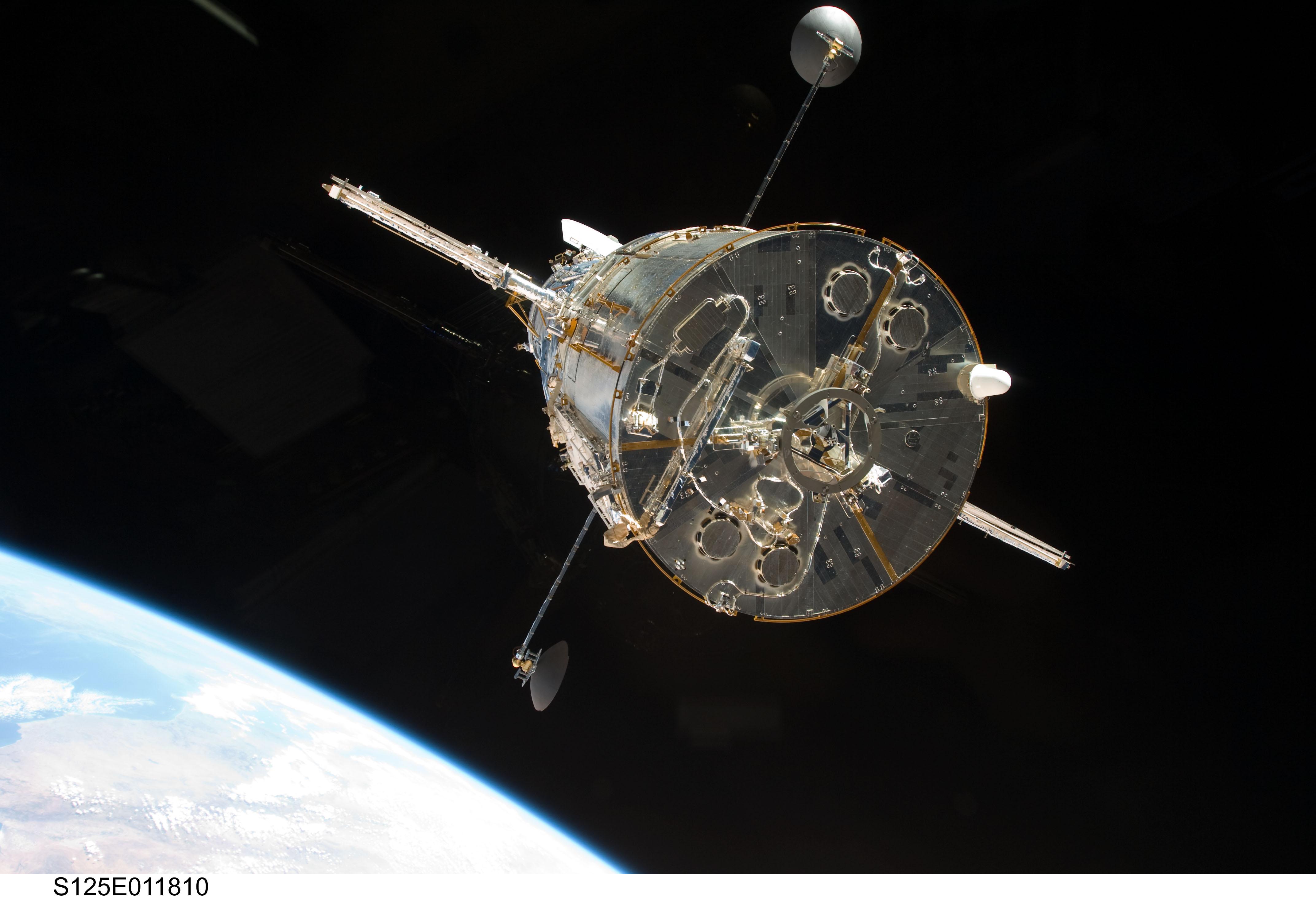}\hfill
    (b) \includegraphics[width=.28\textwidth]{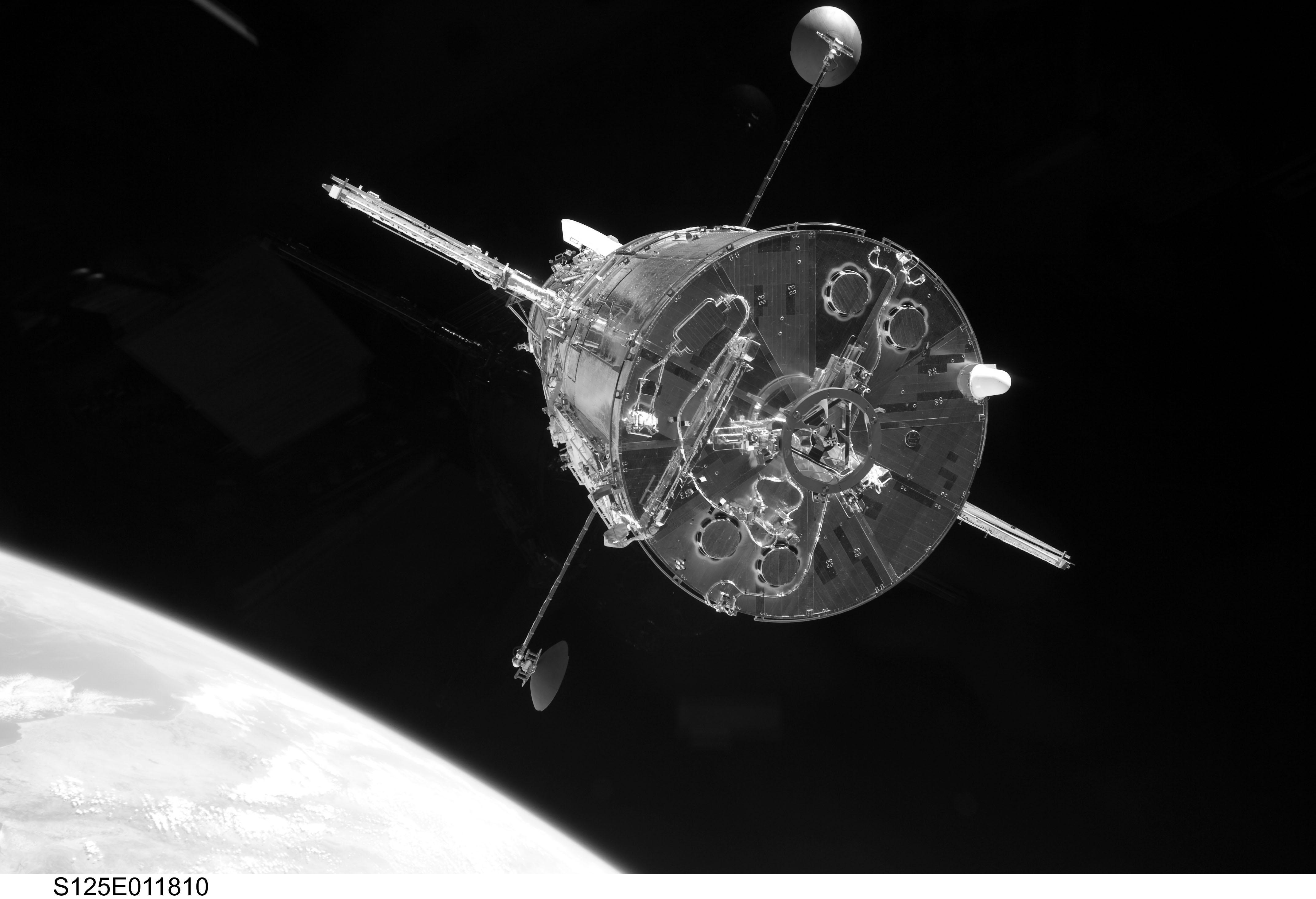}\hfill
    (c) \includegraphics[width=.28\textwidth]{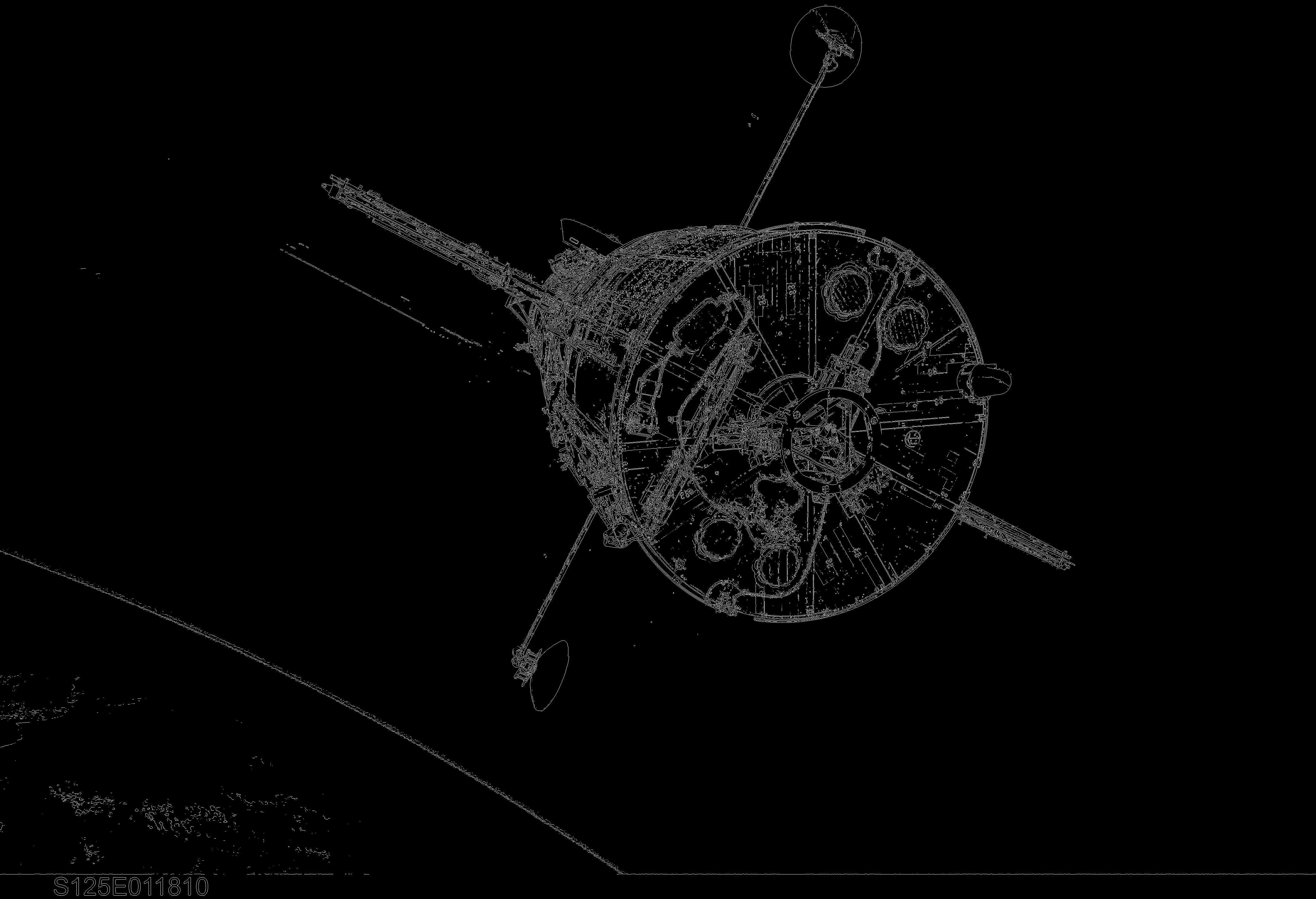}\\ 
    (d) \includegraphics[width=.28\textwidth]{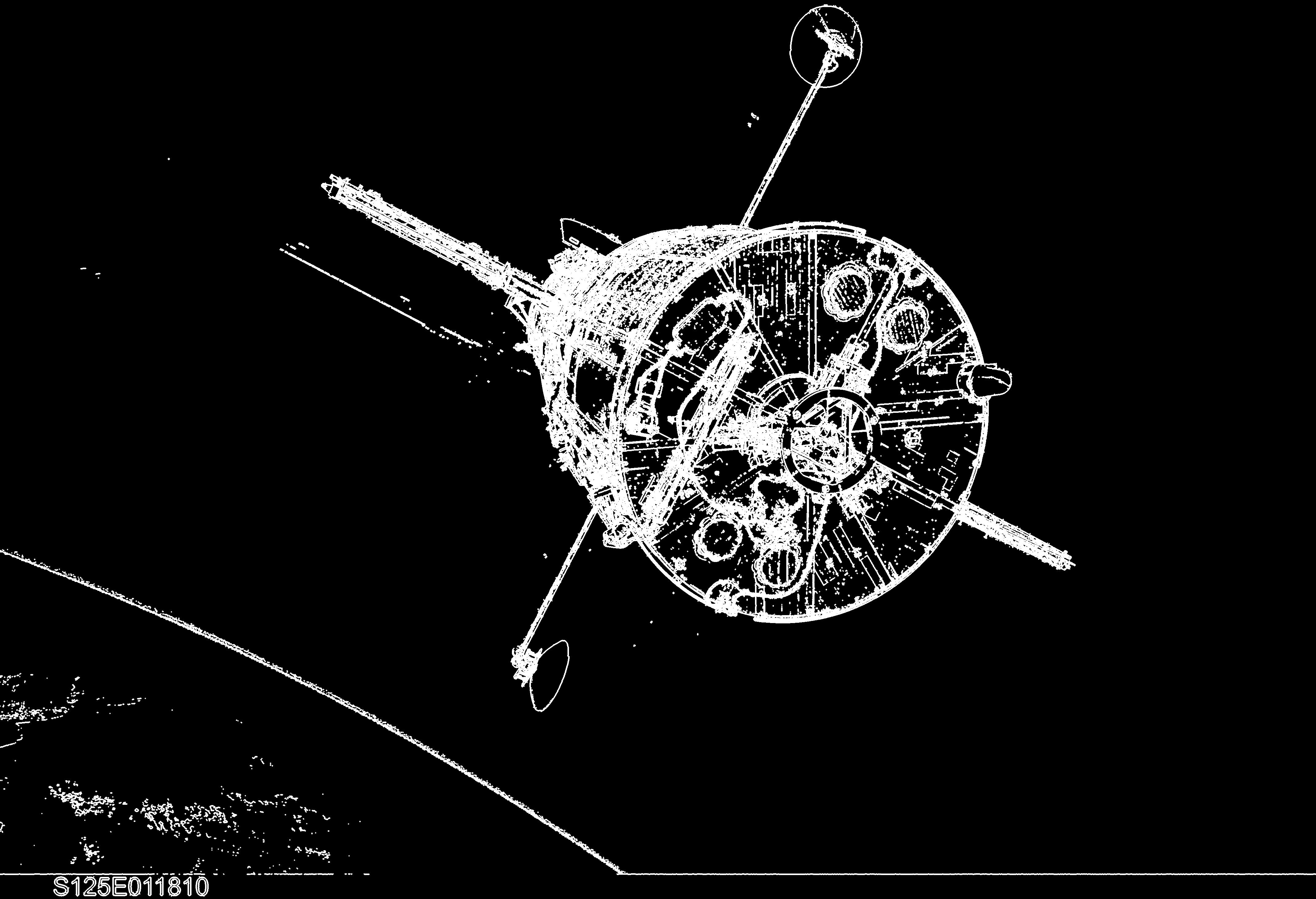}\hfill
    (e) \includegraphics[width=.28\textwidth]{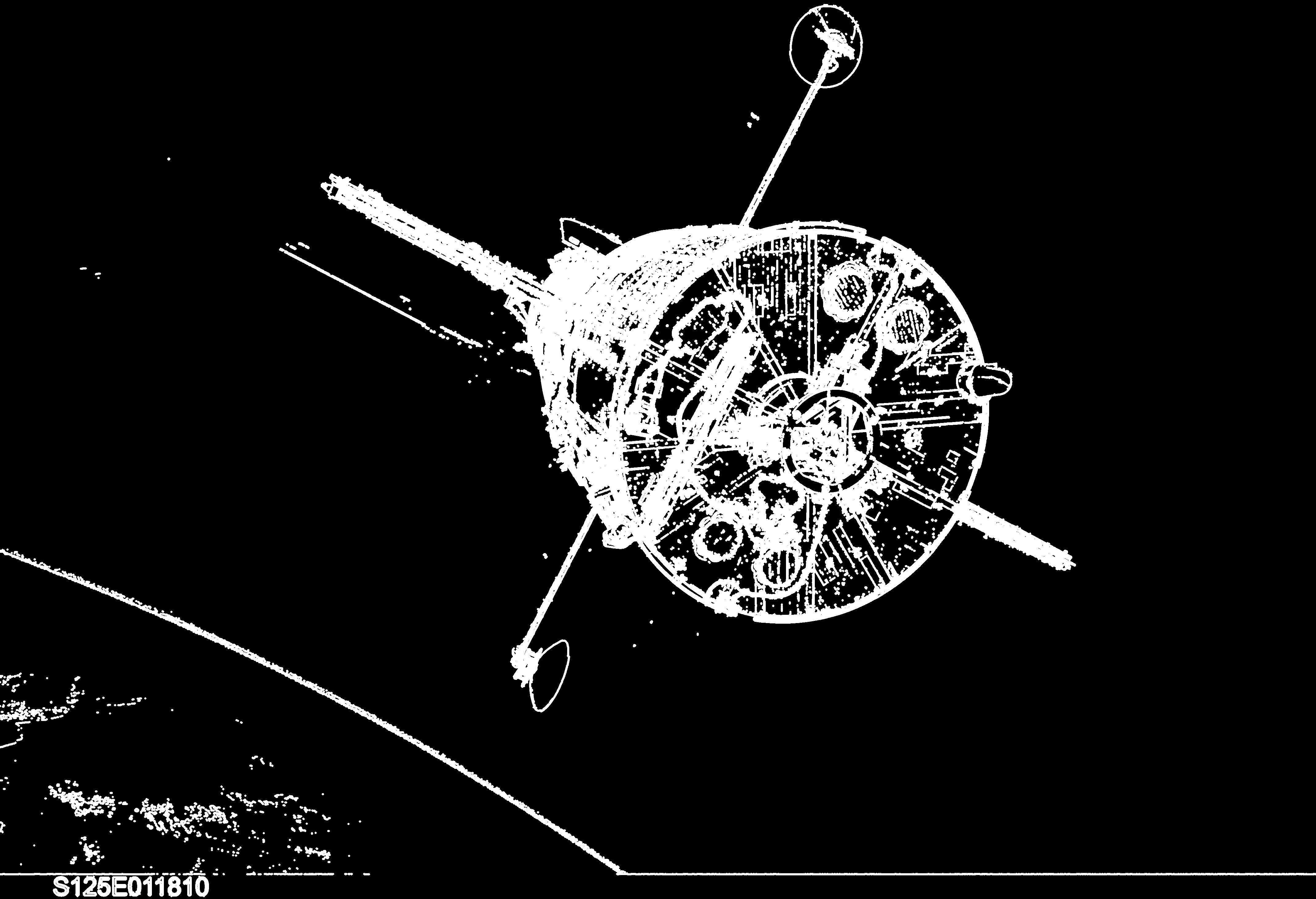}\hfill
    (f)  \includegraphics[width=.28\textwidth]{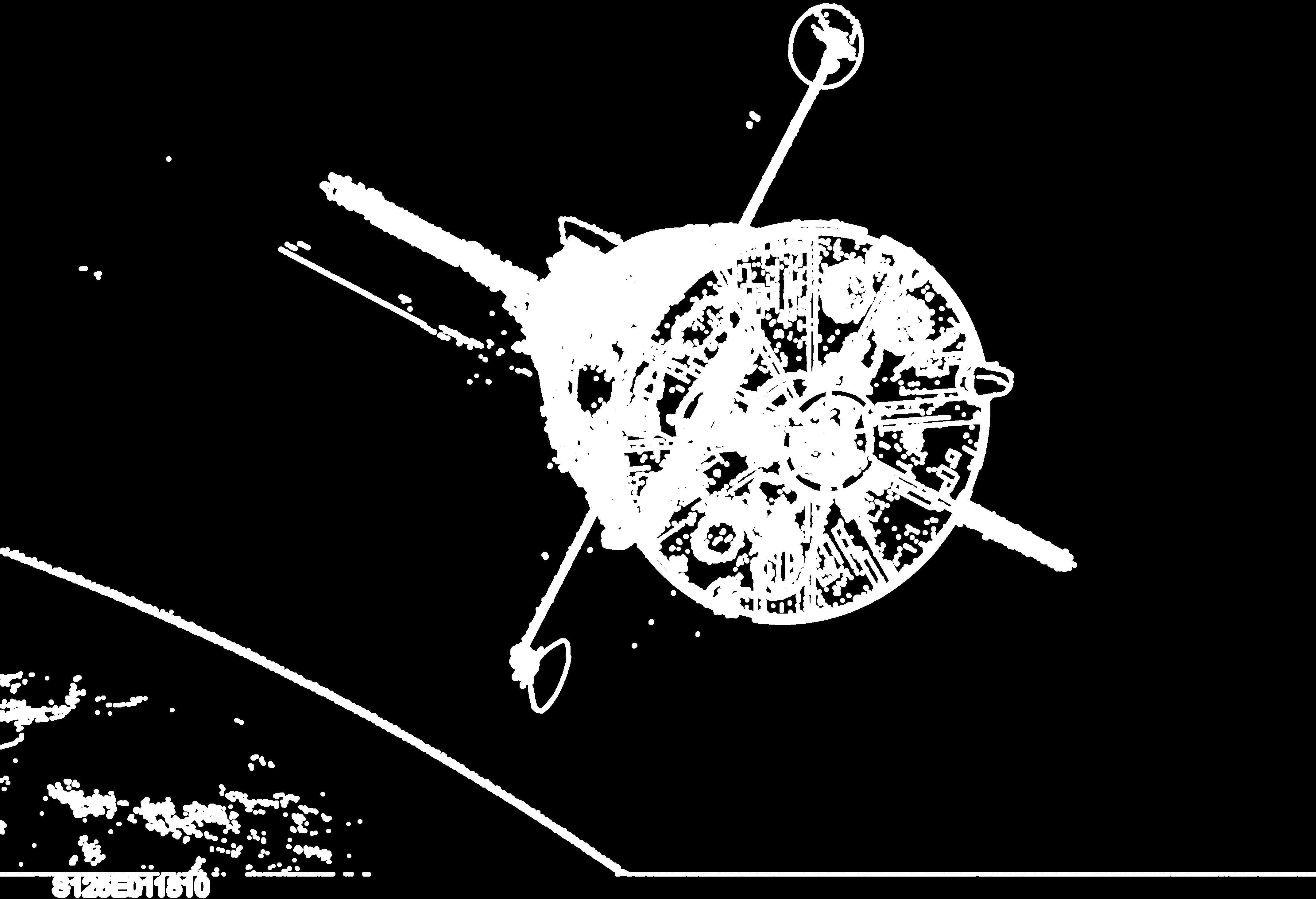}\hfill
\caption{(a) is a color image of the Hubble Telescope, (b) and (c) are a grayscale and black-and-white versions of (a), respectively. Images (c), (d) and (e) are the result of applying a dilation algorithm using three different values for the size of the structuring element.}
 \label{figura2}
\end{figure}
\vspace{0.1cm}
{\bf a) }Exponential speed-up is the result of comparing efficient {\it versus} inefficient algorithms. So, achieving exponential speed-up on a quantum algorithm requires classical algorithms with exponential complexity as counterparts. In classical image processing, the vast majority of problems can be satisfactorily solved with $\mathsf{P}$ algorithms or a polynomial number of  iterations of $\mathsf{P}$ algorithms. This is due to the nature of problems faced in image processing, the finite size of data input and also because success criteria for image processing algorithms is a combination of quantitative outcomes (e.g., a stop criterion as in numerical analysis algorithms) and qualitative results which basically consist of the approval of human beings. So, even for optimization problems in image processing, algorithms do not need to find the global minimum or maximum of the corresponding cost functions, they just need to produce values that make sense to the human eye or that are good enough for machine consumption (like computer vision algorithms, for example).

For instance, let us analyze the behavior of the images and algorithms presented in Fig. \ref{figura2}. The color image presented in Fig. \ref{figura2}.(a) is the Hubble telescope and Fig. \ref{figura2}.(b) as well as Fig. \ref{figura2}.(c) are grayscale and black-and-white versions of (a), respectively.  Producing grayscale and black-and-white digital images is achieved by polynomial-time algorithms.

The borders of image in Fig. \ref{figura2}.(c) are rather weak, i.e., it is hard for the human eye to quickly identify all borders that correspond to the actual Hubble telescope and the different objects that are on its surface. In order to intensify the border contrast in this image, we may use a dilation operator \cite{serramm}, which can be implemented by another polynomial-time algorithm. Images in Fig. \ref{figura2}.(d), Fig. \ref{figura2}.(e), and  Fig. \ref{figura2}.(f) are the result of applying the same dilation algorithm using three different values for one parameter (the size of the structuring element).

Now, the question is: from the second row of Fig. \ref{figura2}, which image is best? The answer will ultimately depend on the preferences of the person in charge of choosing the best image. For instance, we may choose (e) because (d) still has weak borders while  borders in (f) are a bit too thick and we loose some details of the objects that are on the surface of the Hubble Telescope.  This rationale is found over and over in image processing and related fields: algorithms are likely to be found in the $\mathsf{P}$ sphere because of the nature of the problem to solve (in this case, sweeping a mask over the digital image and performing some numerics on each step)  as well as because running the algorithm just a few times will suffice  to produce a result that is satisfactory to the human eye. Indeed, polynomial or exponential algorithmic speed-up must be a goal in QIMP but, in addition to that aim, we should also consider other quality criteria like the suitability of quantum images for human or machine consumption.

\vspace{0.1cm}
{\bf b)} There are two approaches for estimating the complexity of an algorithm: asymptotic analysis and empirical analysis. Asymptotic analysis provides a definite and analytical answer to the amount of computational resources needed to run an algorithm on inputs of arbitrary size. Unfortunately, following this approach is very hard on many algorithms because of the mathematical intricacies faced when calculating asymptotic resource consumption. As an alternative, we may follow the empirical analysis approach: algorithms can be run upon a portfolio of random inputs different in size so that good estimates for typical resource consumption on finite size machines can be produced \cite{7aa17,8}.

Some papers in QIMP have presented algorithms with asymptotic analysis results while some others present numerical results as evidence of algorithmic performance. In both scenarios, quantum algorithms exhibit computational complexities that are roughly of the same order, or polynomially upper-bounded, as those of their classical counterparts.


As an example of asymptotic analysis results, let us present a case study: edge detection in the digital and quantum domains.

Edge detection, a most important activity in image processing and computer vision, consists of identifying sharp intensity changes on an image. The goal of edge detection, in human plain terms, is to identify borders and silhouettes in an image. There are several methods for detecting edges, among them the Sobel operator \cite{sobel01,sobel02} and the Canny operator \cite{canny} methods.

\begin{figure}
   (a) \includegraphics[width=.45\textwidth]{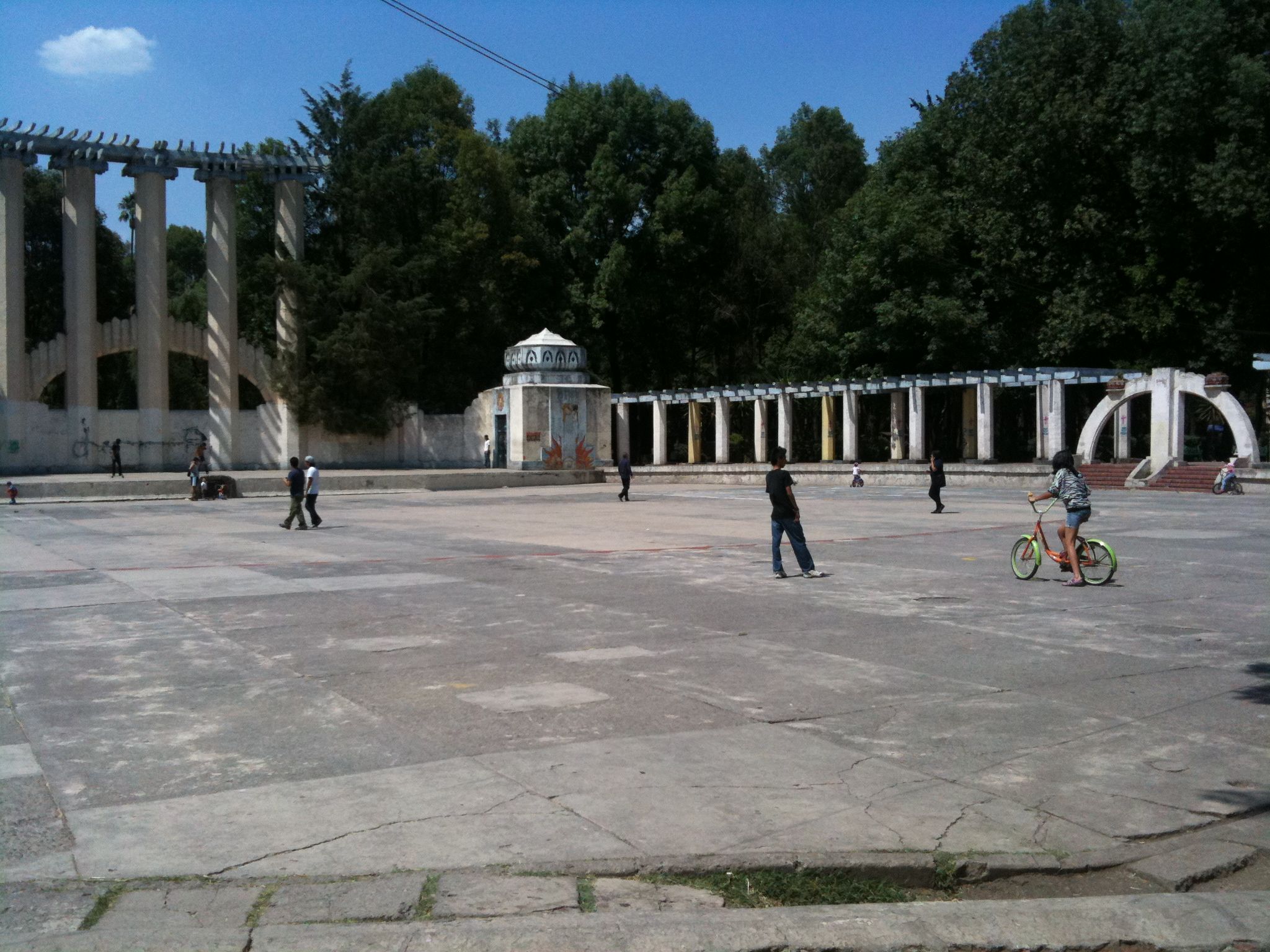} 
    (b) \includegraphics[width=.45\textwidth]{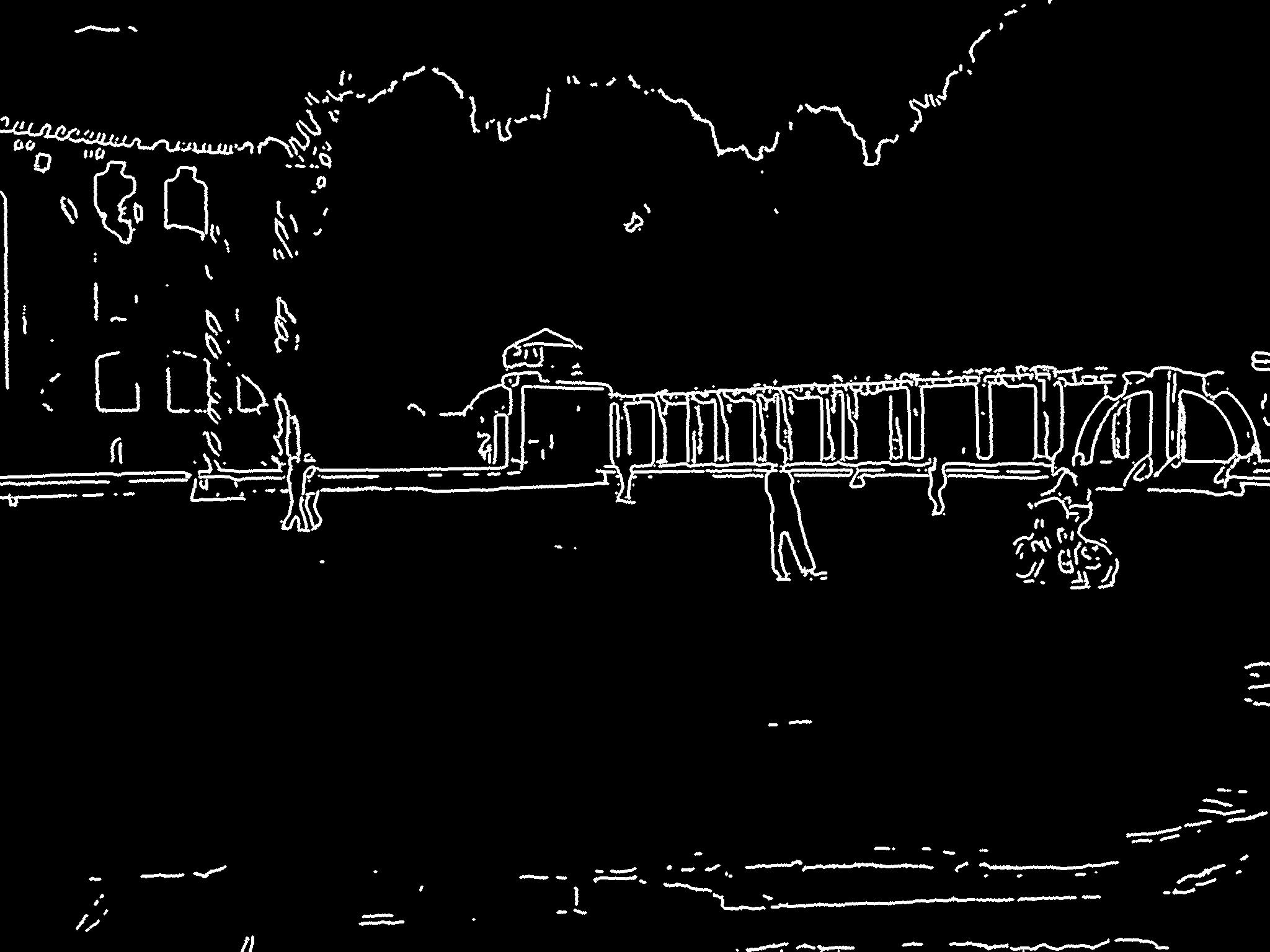}
\caption{(a) is a color image of Parque M\'exico, a famous park in Mexico City and (b) is the result of running the Matlab$^\copyright$  2019 implementation of Sobel algorithm on Fig. \ref{figura3}.(a) (Fig. \ref{figura3}.(b) was previously smoothed using a $4 \times 4$ mask Gaussian filter and its brightness enhanced by running a dilation operator with a radius two disk as structuring element).
}
\label{figura3}
\end{figure}

Fig. \ref{figura3}.(a) is a color image of Parque M\'exico, a famous park in Mexico City. Fig. \ref{figura3}.(b) is the result of running the Matlab$^\copyright$  2019 implementation of Sobel algorithm on Fig. \ref{figura3}.(a) (Fig. \ref{figura3}.(b) was previously smoothed using a $4 \times 4$ mask Gaussian filter and its brightness enhanced by running a dilation operator with a radius two disk as structuring element).

Keeping in mind that the key elementary step in Sobel operator algorithm is the computation of discrete convolution, results for classical and quantum algorithms for computing Sobel operator are:

\vspace{0.2cm}
{\tiny \textbullet} A straightforward implementation of Sobel operator on an image composed of $n^2$ pixels using a textbook definition for computing discrete convolution results in an algorithm of order  $O(n^2)$,  while a more refined implementation based on the Fast Fourier Transform results in an algorithm of order $O(n\log n)$.

{\tiny \textbullet} In \cite{sobelq}, a quantum algorithm for computing Sobel operator on an FRQI image is presented. The complexity of this quantum algorithm, taking into account quantum measurements for extracting the Sobel edge values, is of order $O(n^2)$ on the number of quantum gates, at best (the complexity comparison made in \cite{sobelq}, that of number of bits {\it vs} number of qubits, misses the fact that computational complexity measures the number of elementary steps that an algorithm runs on inputs of arbitrary size).

So, both classical and quantum versions of Sobel operator algorithm are equivalent in terms of computational complexity. As it was the case with the original formulation of Sobel operator \cite{sobel01,sobel02}, next steps for QSobel would likely include using the mathematical machinery of quantum information and quantum mechanics to produce enhanced versions of this algorithm (possibly, achieving polynomial speed-up).

As for QIMP algorithms written under the empirical analysis approach, some papers based on numerical evidence contain claims of high efficiency and performance, claims that need to be further analyzed and tested by the scientific community.  A detailed list of those papers can be found in \cite{3}.

Empirical analysis is a powerful tool in contemporary computer science \& engineering as well as in software engineering. This is because it is often the case that the calculation of computational complexity for modern algorithms using the tools of asymptotic analysis is very hard. Moreover, most of the results of computational complexity are based on the notion of stand-alone computation and some celebrated modern algorithms either run natively on distributed systems or the amount of elementary steps required to run those algorithms exceed the typical power of stand-alone computers. A good example of the difficulties faced when estimating computational complexity is Google's PageRank algorithm.

Let us suppose that we have a digraph $ {\mathbf G}(V,E)$ and that we are interested in ranking its nodes based on the importance of each node, i.e., we want an ordered list $(v_1, v_2, ... v_n)$ where $v_i \in V$, $i \in \{1,2, \ldots, n\}$. PageRank is an algorithm developed to provide a quantitative approach to the qualitative notion of node importance in a digraph. The problem for which PageRank was designed was that of ranking the nodes of the WWW. PageRank was presented in 1999 as a technical report  \cite{pagerank01} and the computation of the node rank basically consists of computing an eigenvector of a matrix $A$ whose entrances are  functions of the incoming links and outgoing links of each node \cite{pagerank02,pagerank03}.

The computation of eigenvectors is a very well known mathematical problem since the XIX Century. Many seminal texts, methods and algorithms about this problem have been published since the beginning of the computer era (e.g., \cite{pagerank04,pagerank05,pagerank06}). Several algorithms for computing the eigenvectors of different families of matrices were presented during those years and an abundant numerical evidence of the polynomial complexity nature of the eigenproblem was presented, but we had to wait until 1999 to definitely know that ``the deterministic arithmetic complexity of the eigenproblem for any $n \times n$ matrix $A$ is bounded by $O(n^3)$'' \cite{pagerank07}.

So, the performance of PageRank is linked to the computational complexity of algorithms for solving the matrix eigenproblem, and having only partial theoretical results as well as abundant numerical estimates was enough to provide the scientific grounds and product development of Google's star product. Furthermore, the tremendous size of the WWW makes the computation of PageRank an interesting challenge for distributed algorithms, and the first provable efficient distributed PageRank algorithm was  published as recently as in 2015 \cite{pagerank08}.

Empirical analysis is a tool that was seldom utilized in the beginning of the quantum computing era but its use is steadily increasing. For example, in \cite{pagerank09,pagerank10}, a quantum algorithm for computing PageRank in a quantum network was presented and \cite{pagerank11} shows a comparison between classical and quantum versions of PageRank. These three papers make extensive use of numerics to estimate the performance of corresponding algorithms. Furthermore, other areas of vigorous research like quantum machine learning are also adopting empirical analysis \cite{8}.

The QIMP community should keep on presenting results based on both approaches. As for asymptotic analysis, procedures and results should be more mathematically rigorous and transparent. With respect to empirical analysis and following current trends in scientific repeatability, testing code and data should be made available to the scientific community. These enhancements will depend on both researchers and the requirements imposed by scientific journals.

{\bf 3. The road ahead - a proposal of future steps for QIMP. }


In addition to the proposals presented along this paper, we put forward the following analysis. Although the development of QIMP technologies that are fully competitive with corresponding digital technologies is highly desirable, future research efforts must avoid attempting to realize quantum versions for every digital image processing algorithm as not all digital image processing algorithms may be appropriate for implementation in the quantum computing realm.  We should choose and develop only those algorithms that either have a better performance than classical algorithms or that significantly enhance the overall performance of image processing tasks via complementing classical algorithms. Furthermore, research must be devoted toward ensuring that formal comparisons can be made between digital algorithms and their corresponding quantum algorithms. This requires the development of metrics for evaluating and comparing the computational complexity of quantum and classical algorithms used for encoding, processing, and recovering images, as well as the physical (both quantum and digital) resources employed for these tasks. In addition, researchers in the field of QIMP should be critical of published results when the theories or methods employed during analysis are controversial.

\newpage{}
QIMP is a component of a greater goal for quantum technology, that of creating a complete integrated quantum technology ecosystem in which data produced by quantum sensors is transmitted via quantum channels. Thus, efforts could be focused on developing quantum algorithms for solving sophisticated operations like image reconstruction, super-resolution, and semantic analysis with quantum image representations, as well as developing potential applications for solving open problems in science and engineering, e.g., machine vision, remote sensing, and health informatics.

In addition, we must note that the development of physical equipment for capturing and processing quantum images is key for making QIMP as pervasive a field as digital image processing. However, the development of such equipment, such as the physical realization of interfaces connecting digital and quantum images (i.e., preparation and measurement), remains a pending task. A crucial aspect of these efforts involves capitalizing on the interdisciplinary nature of QIMP, which spans through physics, optics, computer science, and electrical engineering, to design dedicated hardware either as stand-alone units or as part of larger hybrid systems.

Finally, the development of specific-purpose hardware is a promising approach that conforms well with the contemporary industry-university cooperative research paradigm. For example, analogous to the conversion of analog signals to digital signals using analog-to-digital converters, a very practical effort would be to devise digital-to-quantum converters to integrate the digital and quantum computing realms seamlessly. In addition, we note that most existing QIMP experiments are based on simulations using mathematical software. This can be expected to continue in the future regardless of the availability of hardware owing to the great benefits of computer simulation in the hardware design process. However, such expediencies confuse the implementation of QIMP and digital image processing algorithms thereby restricting the anticipated power of quantum information technologies. Therefore, quantum computing software must be developed to complement the design of future QIMP hardware and facilitate the objective validation of QIMP algorithms (e.g., \cite{qiskitqimp}). Furthermore, quantum error correction must be considered throughout the development process to guard from errors associated with decoherence and other quantum noise. This is another critical area requiring further study.

Feynman's famous lecture title, \lq \lq There is plenty of room at the bottom'', has been an inspiration for many of us, members of the QIMP community, to work towards the development of a branch of quantum science and engineering focused on storing, processing and retrieving visual information using quantum systems, with the higher goal of contributing towards the development of quantum technology ecosystems. This perspective paper has been written with that purpose in mind.

\begin{acknowledgements}
This work was supported by Tecnologico de Monterrey and CONACyT (SNI No. 41594, Fronteras Ciencia 1007).
\end{acknowledgements}

%
%


%
%

\end{document}